\documentclass{vgtc}                          % final (conference style)
\ifpdf%                                % if we use pdflatex
  \pdfoutput=1\relax                   % create PDFs from pdfLaTeX
  \usepackage{graphicx}                % allow us to embed graphics files
  \DeclareGraphicsExtensions{.pdf,.png,.jpg,.jpeg} % for pdflatex we expect .pdf, .png, or .jpg files
\else%                                 % else we use pure latex
  \usepackage{graphicx}                % allow us to embed graphics files
  \DeclareGraphicsExtensions{.eps}     % for pure latex we expect eps files
\fi%

\graphicspath{{figures/}{pictures/}{images/}{./}} % where to search for the images

\usepackage{microtype}                 % use micro-typography (slightly more compact, better to read)
\PassOptionsToPackage{warn}{textcomp}  % to address font issues with \textrightarrow
\usepackage{textcomp}                  % use better special symbols
\usepackage{mathptmx}                  % use matching math font
\usepackage{times}                     % we use Times as the main font
\usepackage{cite}                      % needed to automatically sort the references
\usepackage{tabu}                      % only used for the table example
\usepackage{booktabs}                  % only used for the table example
\usepackage{array}
\newcolumntype{L}[1]{>{\raggedright\let\newline\\\arraybackslash\hspace{0pt}}m{#1}}
\newcolumntype{C}[1]{>{\centering\let\newline\\\arraybackslash\hspace{0pt}}m{#1}}
\newcolumntype{R}[1]{>{\raggedleft\let\newline\\\arraybackslash\hspace{0pt}}m{#1}}
\usepackage{enumitem}

\onlineid{0}

\vgtccategory{Research}

\vgtcinsertpkg

\title{Clear Visual Separation of Temporal Event Sequences}

\author{Andreas Mathisen \thanks{e-mail: am@cs.au.dk}
 \and Kaj Gr\o nb\ae k \thanks{e-mail:kgronbak@cs.au.dk}}
 \affiliation{\scriptsize Department of Computer Science, Aarhus University, Denmark}

\abstract{
Extracting and visualizing informative insights from temporal event sequences becomes increasingly difficult when data volume and variety increase. Besides dealing with high event type cardinality and many distinct sequences, it can be difficult to tell whether it is appropriate to combine multiple events into one or utilize additional information about event attributes. Existing approaches often make use of frequent sequential patterns extracted from the dataset, however, these patterns are limited in terms of interpretability and utility. In addition, it is difficult to assess the role of absolute and relative time when using pattern mining techniques. 

In this paper, we present methods that addresses these challenges by automatically learning composite events which enables better aggregation of multiple event sequences. By leveraging event sequence outcomes, we present appropriate linked visualizations that allow domain experts to identify critical flows, to assess validity and to understand the role of time.
Furthermore, we explore information gain and visual complexity metrics to identify the most relevant visual patterns. We compare composite event learning with two approaches for extracting event patterns using real world company event data from an ongoing project with the Danish Business Authority.
}

\CCScatlist{ 
Human-centered computing -- Visual analytics
}

\begin{document}



\maketitle

%% \section{Introduction} %for journal use above \firstsection{..} instead
\section{Introduction} \label{intro}
\begin{figure}[!ht]
	\centering
	\includegraphics[width=0.48\textwidth]{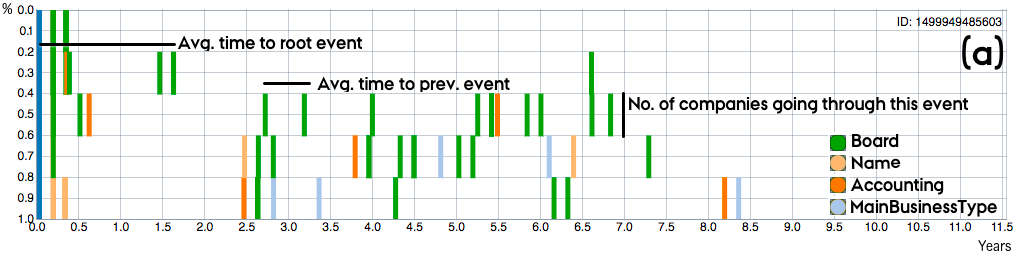}
 	\includegraphics[width=0.48\textwidth]{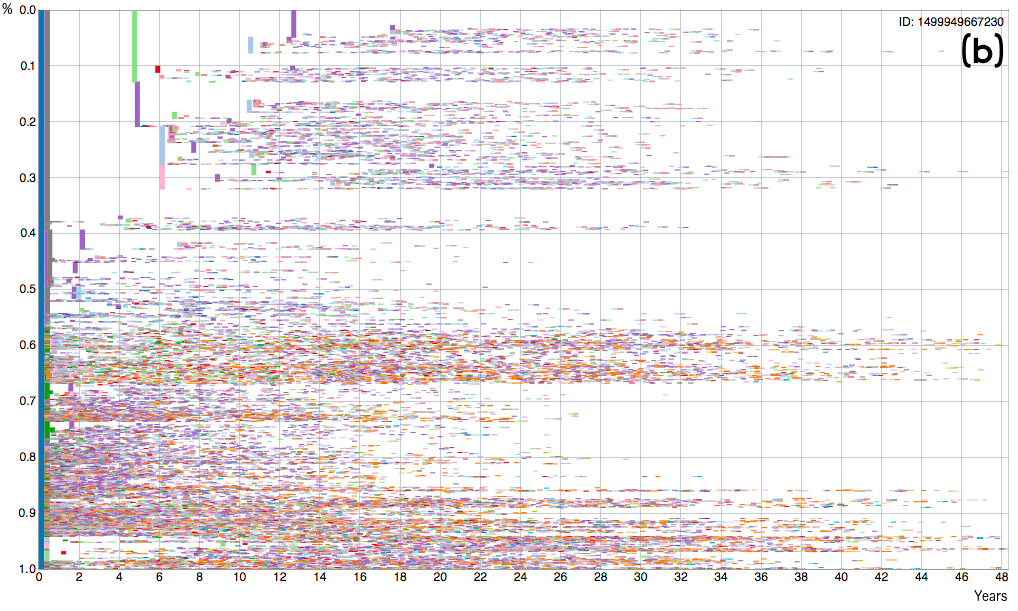}
%	\caption{Event tree with no simplification.}
%	\label{fig:viz2}
%\end{figure}
	
%\begin{figure}[!ht]
%    \centering
 	\includegraphics[width=0.48\textwidth]{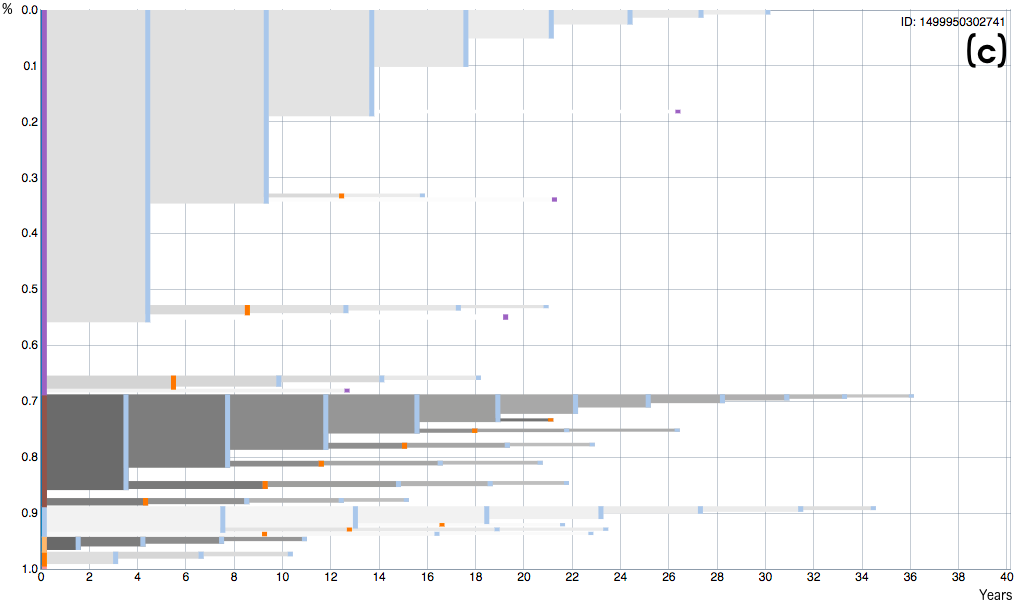}
	\caption{An event tree with few event types and few event sequences (a) that shows how to read the visualization. The event tree in (b) shows a large event sequence collection without any simplification and the event tree in (c) shows an event tree of the same dataset using composite events and outcome percentages encoded in the transitions between events. The y-axis is percentage of the total number of sequences and the x-axis is time.}
	\label{fig:viz}
\end{figure}

Gaining informative insights from temporal event sequences is a challenging task in many real world domains. As shown in Figure \ref{fig:viz}(b), data volume and variety render most sequences unique as evident in a large collection of company event data. While the analysis of temporal event sequences is well-studied within both the visualization and data mining communities \cite{han2007frequent, perer2014frequence, kwon2016peekquence, liu2017patterns} many challenges persist. Existing visualization techniques are often inadequate without appropriate aggregation of the data, since simply visualizing multiple raw event sequences will not provide interpretable information. However, defining meaningful aggregations is also difficult as described in recent work by Liu et al. \cite{liu2017coreflow}. A common approach is to extract frequent sequential patterns, but these techniques often yield an overwhelming number of subsequences on real world datasets. This makes the results unsuitable for manual inspection and it can be difficult to assess the relevance of a derived pattern when inspecting it in isolation. Liu et al. \cite{liu2017coreflow} therefore propose an algorithm to compute branching patterns that describe the most common event sequence flows. This automatic search shares a similar goal to the manual approach of Monroe et al. \cite{monroe2013temporal}, where a series of user-specified simplifications lets the analyst arrive at a simpler representation of the major flows in the data. While the manual approach does not scale very well, the automatic search suffer in terms of interpretability, since it is hard to assess the quality of the derived patterns. Previous work \cite{gotz2014decisionflow, wongsuphasawat2011outflow} have therefore shown promising results when using outcomes to analyze temporal event sequences. In this work we also use sequence outcomes to both reason about pattern quality and to define visualizations that allows the analyst to identify critical flows and to assess pattern quality. To efficiently use temporal event sequences in decision making processes, it is important to leverage both data mining and visualization techniques.  

The motivating use cases for the work presented in this paper comes from an ongoing project with the Danish Business Authority (DBA), which maintains  historical  registration  data  and  financial  statements for more than 1.5 million danish companies. The registration data is essentially temporal event sequences containing events such as changes to business type, name, accountants and board members. The domain experts at the DBA are not data scientists and therefore they lack automated tools to systematically gain insights from their data. Currently their search for suspicious behavior is started from previous cases, hence they often have outcome labels for a subset of the companies, i.e. outcomes describing bankruptcy or fraudulent behavior. Using our methods, the aggregation in Figure \ref{fig:viz}(b) is automatically transformed to the arguably better aggregation in Figure \ref{fig:viz}(c). Analysts can now separate the critical flows (dark) from the less critical flows (light) w.r.t. to the outcome and get insights into how different events affects the outcome probability. 

%The methods presented in this paper is driven by such use cases, where an outcome can be defined for a subset of the sequences. 
%This can make the search for relevant patterns more focused and analysts can afterwards inspect the derived patterns in light of the provided outcomes. 
% Appropriate visualizations are important to allow the analysts to understand the derived patterns, but also to assess how well a given pattern is correlated with the outcomes, i.e. outcomes provide a way to determine pattern quality. 

%Note, an outcome can simply be a specific event occurrence that is interesting to the analyst. 

Pattern mining algorithms are often ill suited when the order in which certain event types occur is irrelevant. In such scenarios, it can be advantageous to group events into a single event prior to the application of any aggregation algorithm. In our work, we automatically group events into higher level composite events since this drastically reduces the number of unique event sequences and thereby enables the simpler aggregation in Figure \ref{fig:viz}(c). Our approach is inspired from the work of Du et al. \cite{du2016coping}, who recently presented a series of strategies for sharpening analytic focus when analyzing temporal event sequences. In this paper, we use clustering to find similar event sequence windows and thereby replace the old event types with new composite events.

% Among others, they mention that it can be beneficial to group events based on time windows and replace all events within a given window with, e.g., the most frequent event type. 

%The transformed event sequences can then either be used in other pattern mining algorithms or aggregated directly. This approach provides a high level description of the data and when combined with sequence outcomes it can hint potential critical flows. 

The event tree visualization (Figure \ref{fig:viz}) is based on the visualization presented in \cite{monroe2013temporal, wongsuphasawat2011lifeflow}. However, given sequence outcomes, we also augment the various paths with outcome probabilities, which allows the analyst to assess the relevance of a pattern.
Furthermore, we empirically validate our approach using information gain and visual complexity metrics. These metrics can also be used to automatically search for interesting views since they describe visual separation w.r.t. to the outcomes.
%This score provides a way to optimize pattern mining algorithms in a way that directly conform with an informative visualization. 
We compare our approach with two versions of the branching pattern mining technique presented in \cite{liu2017coreflow} - the original version and a modified version where the outcome is used to rank the events. Finally, we propose several future research directions for %giving examples of visualizations that can aid 
supporting the understanding and assessment of patterns derived from temporal event sequences.

% Show problem of visualizing everything raw
% problem of sequential pattern mining techniques
% problem with branching pattern mining techniques
% -> importance in transforming the sequences

\section{Related Work}
Visualizing and retrieving insights from temporal event sequences is not a new research focus due do its various application domains such as electronic health records \cite{wongsuphasawat2011lifeflow, monroe2013temporal, gotz2014decisionflow}, different types of clickstreams \cite{lam2007session, liu2017patterns} and company behavior, as presented in this paper. More domains will likely surface in the future as the methods to analyze event sequences improve. Currently, the major challenge in utilizing temporal event sequences for decision making is to cope with volume and variety, which is especially difficult to cope with when the data does not fit into standard tables formats as used in most machine learning scenarios. 

For these reasons, Wongsuphasawat et al. \cite{wongsuphasawat2011lifeflow} and Monroe et al. \cite{monroe2013temporal} have explored event sequence simplification based on user specified queries together with a custom visualization based on the icicle plot \cite{kruskal1983icicle} to display common event paths and average timespans. Zhao et al. \cite{zhao2015matrixwave} proposes a matrix based visualization organized in a wave to provide overview of different web traffic patterns. These methods show promising results when the number of unique event types are manageable to an analyst. In our work, we use the same visualization as in \cite{wongsuphasawat2011lifeflow} to display common event paths. 

Several alternatives to user driven simplifications have been proposed to support the analysis of temporal event sequences. Prominent among these methods is frequent sequential pattern mining \cite{perer2014frequence, kwon2016peekquence, liu2017patterns}, where common subpatterns with a minimum level of support are extracted and visualized to the user. However, Liu et al. \cite{liu2017coreflow} recently pointed out several limitations with these approaches, including limited interpretability and utility. Another approach is to cluster common sequences \cite{wei2012visual, wang2016unsupervised} or search for similar sequences based on a reference sequence \cite{du2016eventaction}. While complete unsupervised clustering of entire sequences have shown promise in certain domains, it can be difficult to assess the relevance of a clustering result and unimportant events might have an overly large impact. 

In the work by Du et al.\cite{du2016eventaction}, the goal is to arrive at a given outcome by identifying appropriate next actions. In general, sequences outcomes are a powerful tool for narrowing down relevant patterns and provide context to the user. In the work by Gotz et al. \cite{gotz2014decisionflow}, users continuously pick event milestones and review how they affect the sequence outcome in the visualization. In our work, we use a similar approach to visualize the outcome probabilities within the different event flows. In the work by Wongsuphasawat et al. \cite{wongsuphasawat2011outflow} event sequences are summarized to provide all possible paths to a given outcome and Lam et al. \cite{lam2007session} use session outcomes to provide the user with context. Related to outcome analysis is cohort comparison and in many cases the two problems can be modelled in the same way. Krause et al. \cite{krause2016supporting} provides a visual tool to efficiently extract relevant cohorts from a database. Malik et al. \cite{malik2015cohort} proposes a system where multiple statistical significance scores for all subpatterns of a certain length are computed and visualized to the user, but their approach does not provide overview of multiple flows.

Event sequences are especially difficult to analyze when events are not only ordered but also timestamped, since it can be a complex task to assess the role of time. Numerous approaches have been proposed to visualize events over time, e.g., in the form of cloudlines \cite{krstajic2011cloudlines}, temporal summaries \cite{wang2009temporal} or visualizations to explore timespans in known processes \cite{loorak2016timespan}. In our work, time is mainly used to reason about which events are close enough to be combined into a composite event. Furthermore, Wongsuphasawat et al. \cite{wongsuphasawat2012querying} and Monroe et al. \cite{monroe2013challenges} also report on challenges with similarity measures for querying event sequences and specifying time interval and absences queries.  

Liu et al. \cite{liu2017coreflow} recently presented the notion of branching patterns (CoreFlow), which are extracted patterns that describe all event sequence flows instead of only frequent subsets of the event sequences. In our work, the goal is also to extract relevant branching patterns w.r.t. to sequences outcome. For this reason, we present and investigate a modified version of the algorithm presented in  \cite{liu2017coreflow} that uses information gain, traditionally used to infer decision trees \cite{quinlan1986induction}, to reason about appropriate milestones, i.e. to generate an event decision tree. The composite event learning method is inspired from the work of Du et al. \cite{du2016coping}, who recently presented a series of strategies for sharpening analytic focus when analyzing temporal event sequences. Among others, they mention that it can be beneficial to group events based on time windows and replace all events within a given window with, e.g., the most frequent event type.

\section{Company Investigation}
The motivating domain for the methods presented in this paper is company investigation in collaboration with the DBA. While the methods we present generalizes to other types of temporal event sequences, we will in this section briefly describe this domain in order to provide intuition prior to the technical details. The DBA maintains several types of data about more than 1.5 million companies in Denmark. This include a registration database, where companies report changes to important information about their business. This include changes in important relations, like accountants and board members, and changes to basic company information, like business type and name. In this work, we model the registration database as temporal event sequences, since all changes are timestamped. In total the database contains more then 50 million events. Removing all sequence information gives little insight, since looking at a single change, e.g., in the board, can both be a positive or a negative change. However, the order of events is sometimes also irrelevant in shorter time windows while major time gaps are still informative. Furthermore, the quality of the data is unknown, since certain changes are not mandatory to report immediately, which means sometimes the order of events introduce more noise than clarity. For these reasons, we introduce the concept of composite event learning in the following section.

%, which can be useful when event types make little sense in isolation or when event sequences are noisy, which is often the case in real world scenarios. 

Domain experts at the DBA usually start their investigations based on knowledge from previous cases, hence relevant labels can be defined for a subset of the companies. These labels include first of all whether a company went bankrupt or otherwise have been forced to close, but also different types of fraud. In this paper, we model this information as a sequence outcome, which will be the driving factor in the analysis scenario we are trying to support. 
%In the following sections we will give concrete definitions of the different terms we use, a method to learn composite events as well as three different ways to compute an aggregated event tree that is suitable for visualization. 
The interactive system we designed, enables domain experts to efficiently visualize relevant event flows w.r.t. to outcome in large sequence collections. For experimental purposes we only use event data prior to 2014 in the evaluation in order to reason about future outcomes, i.e. which companies will close in the period of 2014-2017. Such experiments will also be important in future use scenarios, when domain experts are using the system for decision making.

\section{Learning Composite Events} \label{composite}
In this section, we define the notion of composite events and describe a concept to automatically learn these. As described previously, the main motivation behind generating composite events is to reduce the variety of temporal event sequences which makes it possible to aggregate sequences that would otherwise be unique. The idea behind composite events is therefore to find collections of similar subsequences and replace these with new high-level events. An example from the DBA data is the beginning of a new company. Since new companies usually go through very similar initial processes, their beginning can often be replaced with a single event, which, e.g., includes the addition of several board members and accountants as well as updates to core company information like name and main business type. This furthermore allows for the identification of different types of beginnings where, e.g., there size of the initial board can be important. The following notions will be used throughout the paper:
\begin{description}[style=unboxed,leftmargin=0cm]
\item[Temporal Event Sequence:] A sequence of event tuples $(e_i, t_i)$, where $i$ is the positional index in the sequence, $e_i$ is the event type and $t_i$ is the timestamp. In certain scenarios the sequence contains event triples $(e_i, a_i, t_i)$, where an event also have an attribute $a_i$.
\item[Composite Event:] A grouping of several events from a sequence into a single complex event. 
\end{description}
Note, that in our definition of temporal event sequences, two events from the same sequence are allowed to share the same timestamp. There exist numerous approaches for retrieving common patterns in temporal event sequences that can be used to find composite events including frequent pattern mining \cite{han2007frequent}, temporal abstractions \cite{shahar1997framework} or user-specified find-and-replace methods \cite{monroe2013temporal}. However, not all methods can deal with events that share the same timestamp. In this paper, we use a simple bucketing by time period approach, coined as a strategy to simplify temporal event sequences in \cite{du2016coping}, as the foundation for defining composite events. The sequences are divided into equal time segments that each will constitute a composite event. By counting event type occurrences within each segment (and potentially attribute occurrences) conventional clustering methods can be used to define similar segments and thereby find composite events. Concretely, for window size $w$ and number of clusters $k$ we do the following:
\begin{description} \label{comp_alg}
\item[1. Segmentation:] Divide each temporal event sequence into equal time segments of size $w$. % HB: equal? equally long segments? of equal duration?
\item[2. Feature Generation:] Count event type occurrences in each segment generating a feature for each event type, effectively ignoring sequence order within the window segments.
\item[3. Clustering:] Partition all segments into $k$ groups using the k-means clustering algorithm \cite{arthur2007k}. The $k$ groups constitute the composite event types.
\end{description}
This realization of the three steps in the composite event learning concept
%HB: the concretizatons/concrete instantiations/strategies? for the 3 steps in the  concept/layout/algorithmic framework above 
introduces the challenge of finding the optimal window size $w$ and number of clusters $k$ for some notion of \textit{optimal}, which in our use case is an event tree with good separation of the outcomes and good predictive power beyond the dataset. If an overly large $w$ and a small $k$ is used, the resulting aggregation will potentially be an oversimplification, which can result in an event tree with low separation of the outcomes. On the other hand, if an overly small $w$ and a large $k$ is used, the following aggregation will potentially be overfitted to the dataset and have little predictive power. We will describe sequence aggregation methods and quality metrics in the subsequent section. The exact choice of $w$ and $k$ dependents on the use case. The time between events is usually larger in the company event data compared to, e.g., weblogs or medical records, hence appropriate choices of $w$ are likely also larger for this use case. 

The proposed realization is only one way to find composite events and there exist multiple methods for implementing the three steps.
%HB: term, see above!
%HB concretizing/implemeting
%HB: in the concept/layout/algorithmic framework above 
Dynamic window sizes or the addition of more features, such as average time between events or features based on event attributes, could be immediate extensions of our realization.
%HB: term approach. 
When introducing increasing complexity in the mining approach, 
though, complexity is also introduced in the visual interface, since additional information about feature types and varying window sizes needs to be conveyed to the user. Different clustering algorithms or alternative similarity measures can also be used in step 3. A study on similarity measures for text document clustering using k-means showed that the euclidean distance can be outperformed by alternative measures in this domain \cite{huang2008similarity}. In text document clustering features are usually also frequency-based, hence it seems worthwhile 
to investigate different similarity measures for event frequencies in future work. However, it is not the main focus of this paper to pick the optimal segmentation, feature generation or clustering methods, but rather how to utilize the concept of composite event learning to generate visualizations that provide relevant insights about temporal event sequences.

% The event attributes can for instance be used to generate features or, better clustering methods can be utilized.

\section{Sequence Outcome, Aggregation \& Quality}
In the following, we will describe the three different temporal event sequence aggregation methods investigated in this paper. Furthermore, we will define how sequence outcomes are used to encode flow probabilities and to score the overall pattern. Sequence outcomes are important in many event analysis scenarios and is usually the occurrence of a certain event type. An example is health outcome analysis as described in \cite{gotz2014decisionflow}, where analysts and epidemiologists study data to understand what factors influence certain health outcomes. As previously described, outcome is also the driving factor behind the motivating analysis tasks at the DBA. In this paper, we define an outcome as a \textit{special} event occurrence. 
\begin{description}[style=unboxed,leftmargin=0cm]
\item[Sequence Outcome:] An event sequence outcome is a special event tuple $(o_i, t_i)$, where $i$ is the sequence position, $o_i$ is the outcome type and $t_i$ is the timestamp of the outcome. 
\end{description}
Note, that any event type can therefore be thought of as an outcome and in the DBA case the relevant outcomes include for instance bankruptcy and different types of fraud. Given an outcome, the sequences can be divided into two groups - the sequences that include the outcome and those that do not. While the basic analysis scenario constitute a binary distinction, the methods presented in this paper can be extended to multiclass or numeric outcome scenarios.

\subsection{Sequence Aggregation} \label{aggregation}
We consider three different methods to aggregate temporal event sequences that are suitable for visualization - one method where the simplified sequence collection is aggregated and two methods where descriptive patterns are extracted from the raw event sequences. The first method computes a hierarchical structure of all unique event paths. The second method is the branching pattern algorithm (CoreFlow) proposed by Liu et al. \cite{liu2017coreflow}, which computes the most common flows. The last method is a modified version of the branching pattern algorithm, where we use entropy based information gain using the sequence outcomes to rank the events instead of using the overall frequency. Entropy is often used in decision tree algorithms to greedily choose the best attributes to branch on. Entropy and information gain will be explained in section \ref{quality}. The result of all approaches is a hierarchical tree structure as described in \cite{wongsuphasawat2011lifeflow}.
\begin{description}[style=unboxed,leftmargin=0cm]
\item[Simplified Aggregation (SA):] First, all sequences are simplified by computing composite events as described in Section \ref{composite}. Sequences with the same prefix will follow the same path in the data structure, and when two sequences no longer consist of the same series of events the data structure will branch. The hierarchical structure contains a path for each unique event sequence, hence in cases with high variety there will be little to no aggregation. In our approach we root the aggregation at the first event in each sequence, however, it is also possible to root the aggregation on, e.g., the first occurrence of a certain event type or other user-specified alignment points \cite{monroe2013temporal}. 

\item[Most Common Pattern (MCP, CoreFlow \cite{liu2017coreflow}):] The Rank-Divide-Trim algorithm presented in \cite{liu2017coreflow} recursively ranks all events w.r.t. frequency (the number of sequences an event occurs in), divides the sequences based on the highest ranking event type (where ties are broken by minimum average sequence index) and then trims the sequences up to the first occurrences of the chosen event. This approach creates the same hierarchical tree structure as the full aggregation, however, only including the most frequent occurring events.

\item[Most Separating Pattern (MSP):] This method uses the same rank-divide-trim procedure as for the most common pattern but with a different ranking function. By using entropy-based information gain with the sequence outcomes to rank the events, the resulting hierarchy will include the most separating events w.r.t the outcome. This method is similar to building a decision tree. 
\end{description}
All aggregation methods can be used either with or without the event sequence simplification described in section \ref{composite}. In section \ref{eval}, we will provide a comparative evaluation of the SA method, which uses sequence simplification, and the two pattern extraction methods on the raw event sequences, with data from the DBA use case. The evaluation also serves as an example of how a potential evaluation can be done in other domains in order to find the best combination of event sequence simplification and aggregation method.

\subsection{Quality Metrics} \label{quality}
Sequence outcomes provide a way to compute pattern quality empirically. Since the outcome of interest is often heavily outnumbered in real world datasets, i.e. only a subset of the sequences contain the outcome of interest to the analyst, it can be difficult to argue about pattern quality for two reasons. First, if the dataset is used to predict outcome, you get a very high accuracy simply by repeatedly guessing on the dominant outcome, which in the DBA domain is the same as always predicting no bankruptcy. Second, the goal of an analyst is not only to get a list of the next instances that will have the interesting outcome, but also to investigate which events or combinations of events that have an influence on the outcome in order to limit the number of instances in an investigation and later fuse the gained knowledge with other sources of data. For these reasons, we use entropy-based information gain to measure pattern quality which is commonly used to reason about appropriate splits when building decision trees. Furthermore, we also use two metrics describing visual complexity presented by Monroe et al. \cite{monroe2013temporal}. 

\begin{description}[style=unboxed,leftmargin=0cm]
\item[Entropy:] Entropy is a measure of sample homogeneity. If a sample is completely homogeneous, i.e. all sequences lead to the same outcome, the entropy is zero and if the sample is equally divided, i.e. contains the same number sequences leading to each of the outcomes, the entropy is one. The entropy $E$ of a sample $S$ is calculated as
$$E(S)= - \sum_{i=1}^{c} p_i log_2 p_i$$
where $p_i$ is the probability of outcome $i$.  

\item[Information Gain:] The gain in information is the decrease in entropy after a dataset is split into a number of samples. The information gain $IG$ of a sample divided into the partitioning $A$, where $A$ in our case describes the resulting partitioning given by the extracted event paths, is calculated as
$$IG(S, A) = E(S) - \sum_{v\in A} \frac{|S_v|}{|S|} E(S_v)$$
where $v$ is a unique path to the point of splitting in the event tree. Effectively, we will use IG to reason about the initial outcome distribution versus the outcome distributions of the chosen samples in our event tree. The samples can be defined by choosing a minimum level of support, which we denote as the point of splitting. This can, e.g., be the subgroups just before the samples become smaller than 5\% of the total number of records. This notion is similar to the concept of minimum level of support from frequent sequential pattern mining. If the minimum level of support is simply set to 0\%, the leafs of the event tree will be used as the chosen samples.

\item[Visual Complexity:] We compute visual complexity using two measures proposed by Monroe et al. \cite{monroe2013temporal}. (1) The average height of the vertical elements as percentage of the display height, i.e. the size vertical bars describing the different events in the event tree as percentage of the total number of records. (2) The number of elements in the event tree. These measures builds on two central notions of visual complexity; separability and information density. Few visual elements means low information density and large visual elements are easier to distinguish from each other, hence reduce perceived complexity. 
\end{description}

Both information gain and visual complexity are important metrics. High information gain alone does not necessarily imply good generalization beyond the dataset, since the simple split into raw event sequences, i.e. multiple samples of size 1, will give the maximum information gain. Lowering the visual complexity is therefore important for both interpretability as well as generalization purposes. Visual complexity is similar to the generalization heuristics of decision trees, where smaller trees should in theory generalize better.  

\begin{figure*}[!ht]
	\centering
 	\includegraphics[width=\textwidth]{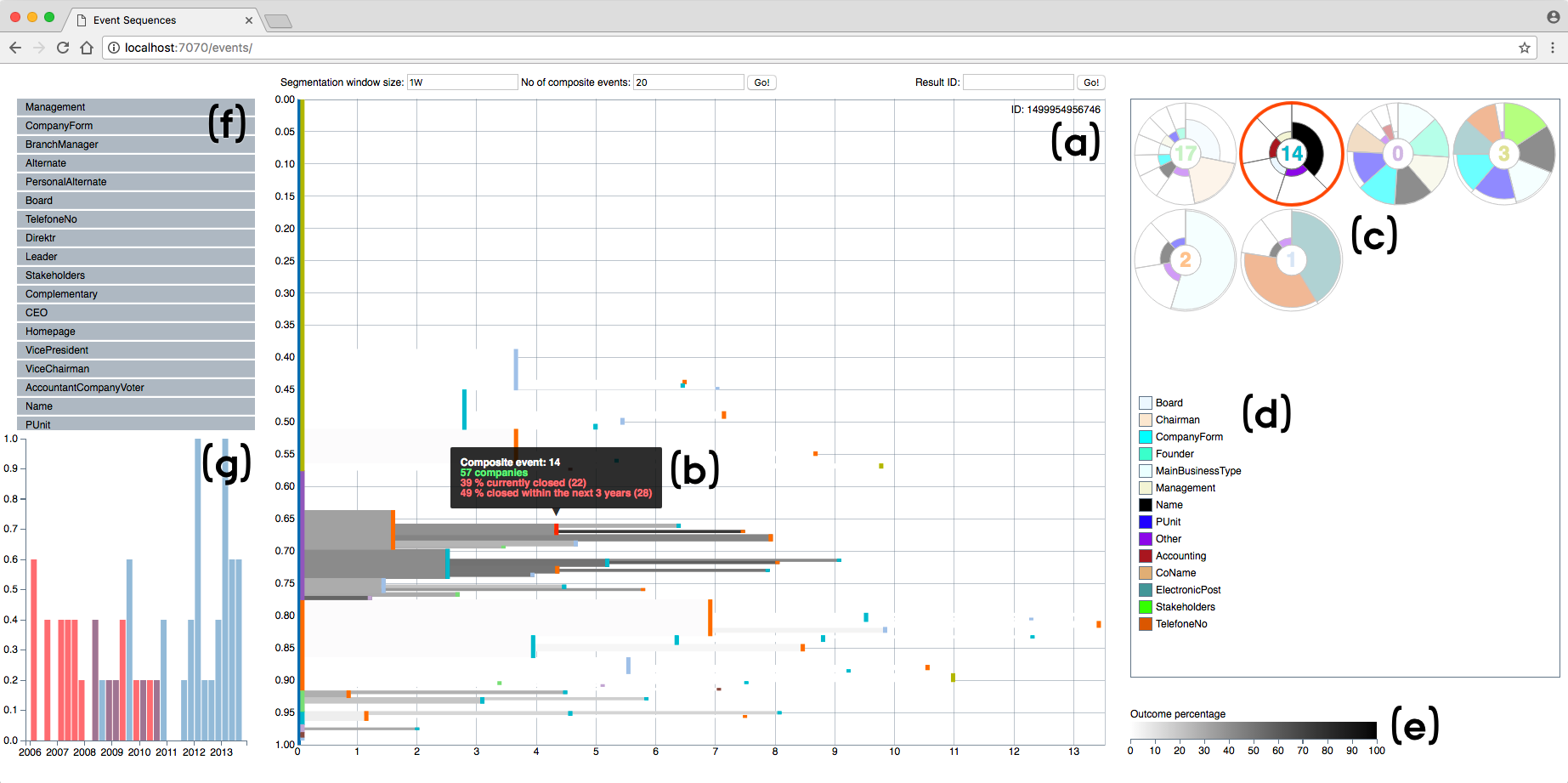}
	\caption{The screenshot shows our prototype tool with dataset 1 (cf. Table \ref{tbl:datasets}). The main view (a) shows the probabilistic event tree with composite events and outcome percentage encoded in the transitions between events --- white means that zero companies have closed and black means that all companies taking this subpath have closed cf. (e). Hovering a composite event in the hierarchy will show the corresponding outcome statistics (b), highlight the composite event in (c) and display a time histogram (g) that shows when the companies went through this event (red are closed and blue are normal). The components of the composite events can be inspected using the aster charts (c), which show the composite events that the user is currently interested in, as well as the corresponding event legend (d). Users can also specify the size of the segmentation window, the number of composite events to compute and limit which event types to use (f) as well as load previous results.}
	\label{fig:viz_comp}
\end{figure*}

\section{Visualizations \& Interactions}
We designed a prototype system that is able to visualize probabilistic event flows of large data collections using any of the aggregation methods described in section \ref{aggregation}. When the SA method is applied, users can currently define the segmentation window size (\textit{w}), the number of composite event types (\textit{k}) (cf. Section \ref{comp_alg}) and limit which raw event types to use (Figure \ref{fig:viz_comp}(f)). While this allows the analyst to compute any desired aggregation, it provides little support in choosing suitable parameters for a given dataset, hence we discuss how this interaction can be improved in Section \ref{future}. The prototype has two central views, one for the event tree, visualizing also outcome probabilities at the event transitions, and one to investigate the composite events. In the following, both views will be described in detail as well as central interactions.

\subsection{Probabilistic Event Tree}
Most tree visualizations can be applied to display aggregated event patterns. In our prototype, we use the hierarchical time visualization presented in \cite{wongsuphasawat2011lifeflow} including also outcome probabilities at each event transition (Figure \ref{fig:viz_comp}(a)). The x-axis represents time, i.e. average time between events and average time to the root, which in this case is the start of the extracted sequence time periods. The y-axis represents percentage of the total number of sequences. Colored vertical elements represent a specific type of event or a composite event, i.e. the large yellow element shows that almost 60\% of the sequences start with composite event 3 and then around 30 \% of those sequences continue with composite event 1, 2 or 14, while the remaining sequences end. The composite events can be inspected using the aster charts (Figure \ref{fig:viz_comp}(c)) which will be described in the following section, i.e. the color of the vertical element corresponds to the color of the composite event number. Relevant composite event descriptors (aster charts) can be added by clicking the composite events in the the event tree. 

Outcome percentages of the individual subpaths are encoded with a continuous black and white scale in the transitions between events, where 0\% is white and 100\% is black (Figure \ref{fig:viz_comp}(e)). This allows the users to quickly identify the relevant subpaths for a given analysis task and see how the outcome probability changes with each event on a path. Furthermore, the event tree visualization includes two simple but powerful interactions. First of all, users can zoom on both axes with mouse scrolling, which allows the analyst to focus on relevant sub paths. Secondly, users are provided with a tooltip when hovering an event bar with basic statistical information about the outcome (Figure \ref{fig:viz_comp}(b)). Currently, we also report on future outcome, i.e. outcome beyond the current data extract used to generate the event tree, which allows the analyst to assess whether the current subpath generalizes. The data extracts will be explained in Section \ref{eval}. We will also discuss how to extend the interaction opportunities such that users can experiment with different time periods to investigate how well certain patterns explain future outcomes. 

When an event in the event tree is hovered, users are also provided with a time histogram (Figure \ref{fig:viz_comp}(g)) that summarizes the timestamps of the events in the chosen subgroup, i.e. when each sequence went through the hovered event in absolute time. Blue summarizes one outcome (in this case companies that are still alive) and red summarizes the other outcome (closed companies) -- the histogram then shows both blue and red bars as two overlays for easy height comparisons. First of all, the time histogram allows the analyst to identify whether certain events occurred at certain points in time and, secondly, it allows the analyst to identify whether the time distribution is different for the two outcomes. In cases where the distribution is different, the time histogram can potentially help the analyst narrow down interesting sequences even further by, e.g., first investigating the blue instances that occur closest to the red instances.

% time histogram

\subsection{Composite Event Inspection}
The components of the composite event types can be inspected using (1) the aster charts (Figure \ref{fig:viz_comp}(c)) and (2) the event legend (Figure \ref{fig:viz_comp}(d)). Each aster chart represents one composite event type, which in our case is a resulting cluster of the k-means algorithm. A slice represents one original event type and the color of the slice matches the event legend. The width of a slice is the feature mean, i.e. average number of event occurrences within each segment, proportional to the sum of all feature means within a cluster. The height of a slice is the feature mean proportional to the means of the same feature in the other clusters. These compact visualizations of the cluster centers are suitable in this scenario since several feature means (event counts) are zero, i.e. those events did not occur within the segments and are therefore irrelevant in order to understand a certain composite event. To make the circle charts even simpler, we combine features with very low means into one category, which can then later be inspected if desired, e.g., to search for outliers. Users can quickly compare multiple composite events for common event types and average number of occurrences by looking at both color, height and width of the slices. As an example, if a slice has a height of 50\% it means that there is a composite event where this type of event occurs twice as frequently. Hovering the aster charts provides the user with statistical information about the individual clusters.

The event legend (Figure \ref{fig:viz_comp}(d)) describes actual event types and is linked with the other views, i.e. it shows only the event types of the composites events that the user is currently investigating. Hovering the event legend provides the user with overall statistical information about the individual event types, which can be used to provide context to the statistical information about the composite events. The composite event inspection is work in progress and we aim to do a user study to evaluate how to optimally visualize these (cf. Section \ref{future}).

\section{Evaluation} \label{eval}
In this section, we will first present a comparative evaluation of the different aggregation methods presented in Section \ref{aggregation} on three different datasets from the DBA using the quality metrics described in Section \ref{quality}. We will also describe a use case example to illustrate the practical implications for domain experts at the DBA. Statistics for the chosen datasets are shown in Table \ref{tbl:datasets}. The datasets represent three different business types (IT-service, Building Entrepreneurs and Retail). Before analyzing the event sequences, several choices have to be made regarding time. In this evaluation scenario, we use only the last 10 years of the event history for each company, since otherwise the event tree would by highly biased by the initial updates of a company. The investigated outcome of this evaluation is \textit{risk of bankruptcy}, which describes that the company have been flagged in the database as either bankrupt, forced to closed or in a pre-state for either of the two. Table \ref{tbl:datasets} also show the overall percentage of companies that have experienced \textit{risk of bankruptcy}. For the companies that do experience \textit{risk of bankruptcy}, we use only the events one year prior to the outcome, since in a real analysis scenario domain experts want to identify companies worth further inspection some time before the outcome actually occurs. Furthermore, we only use data prior to 2014 in general, which allows us to reason about how extracted patterns generalize to the following years. For reference, Table \ref{tbl:datasets} also show the general chance of picking a company that will experience \textit{risk of bankruptcy} after 2014. The different choices regarding dataset preparation means that multiple parameters can be tuned and tested, which we will discuss further in section \ref{future}. The following results have been computed with a one week window size and 25 composite events for the simplified aggregation method.

% Different window sizes and number of clusters

\begin{table}
	\centering
	\begin{tabular}{l | C{1.2cm} C{1.5cm} C{1.2cm}}
	   	& Dataset 1 & Dataset 2 & Dataset 3 \\
		\hline
	    No. of sequences       & 3475 & 4008 & 25122  \\
    	No. of events          & 30494 & 41151 & 222138   \\
	    No. of unique events   & 42 & 59 & 68 \\
	    Business type          & IT-Service & Building Contractors & Retail \\
	    Risk of bankruptcy (\%)   & 8   & 58    & 13 \\
	    Future pred. prec. (\%)& 3.5 & 2.6 & 4
	\end{tabular}
	\caption{Dataset characteristics}
	\label{tbl:datasets}
\end{table}

\subsection{Comparative Evaluation of Aggregation Methods} \label{comp_eval}
\begin{table}
	\centering
	\begin{tabular}{l | c | ccc}
	   	& Dataset & MCP & MSP & SA \\
		\hline
	    Information Gain   &    & 0.003 & 0.151 & 0.153 \\
    	Average Height     & 1  & 6.71  & 5.33  & 2.10  \\
	    Number of Elements &    & 58    & 67    & 36 	\\
	    \hline
	    Information Gain   &    & 0.431 & 0.713 & 0.744 \\
    	Average Height     & 2  & 6.04  & 3.95  & 2.28 \\
	    Number of Elements &    & 69    & 84    & 55 \\
	    \hline
	    Information Gain   &    & 0.037 & 0.262 & 0.246 \\
    	Average Height     & 3  & 6.37  & 4.85  & 2.76 \\
	    Number of Elements &    & 57    & 66    & 37 \\
	\end{tabular}
	\caption{Evaluation results for the three different aggregation methods on the three datasets with a minimum support of 1\%.}
	\label{tbl:info_gain}
\end{table}

Table \ref{tbl:info_gain} shows the evaluation results for the three different aggregation methods on the three datasets with a minimum support of 1\% for the subgroups. The MCP method provides the least information gain w.r.t. the outcome, which is not surprising since there is no guarantee that the most common event flows will also describe relevant flows w.r.t. outcome. However, the MCP method is the best for reducing visual complexity when measured using average height of the visual elements. The MSP and SA methods provide similar information gains w.r.t. the outcome despite that the MSP method is using the outcome to generate the aggregation and the SA method is complete unsupervised. The SA method computes the fewest visual elements in the event tree. However, these elements are composite events that need further investigation and the SA method is therefore not necessarily superior in terms of visual complexity, especially since the MSP method computes larger visual elements. While the MCP method is superior in terms of visual complexity, the method does not let the analyst gain insights into how the outcome flows differ. Since the MSP and SA methods provide similar information gain w.r.t. the outcome, we will in the next section also explore how an event tree can assist the analyst narrow the search for companies that will experience the negative outcome in the future. 

\subsection{Use Case Example} \label{use_case}
In this section, we will describe insights an analyst can get when using the prototype tool with the SA method and dataset 1 as presented in Figure \ref{fig:viz_comp}. First of all, the event tree shows that several companies perform very few updates in the DBA database. However, it is not simply the well-functioning companies that update their information since there exist both dark and light paths. Some of the most common composite events are 1, 2 and 14, which mainly consist of updates to business type, name and contact information. Furthermore, the two most common sequence beginnings are the composite events 3 and 0, which describes the start of a very light path and the start of a very dark path, respectively. The event tree allows an analyst to efficiently identify the most common negative and positive paths. By zooming and hovering the event tree and the composite event glyphs, the analyst can inspect exactly what the composite events are made of and identify related composite events as well as relevant subpaths. The time histogram shows that a lot of updates happens around 2008, which we later found out is because the DBA introduced new regulations at that time. 

\begin{table}
	\centering
	\begin{tabular}{l | c | ccc}
	   	& Subgroup & MCP & MSP & SA \\
		\hline
	    No. of sequences        &   & 469   & 745   & 617 \\
	    Risk of bankruptcy (\%)    & 1 & 34    & 30    & 30  \\
	    Future pred. prec. (\%) &   & 12    & 13.9  & 17.2 	\\
	    \hline
	    No of sequences         &   & 3994  & 2576  & 2300 \\
	    Risk of bankruptcy (\%)    & 2 & 58    & 88    & 95  \\
	    Future pred. prec. (\%) &   & 2.6   & 7.6   & 10.3 \\
	    \hline
	    No of sequences         &   & 5544  & 7264 & 6425 \\
	    Risk of bankruptcy (\%)    & 3 & 43    & 42   & 45 \\
	    Future pred. prec. (\%) &   & 17.2  & 18.9 & 20.5
	\end{tabular}
	\caption{Statistics when looking at subgroups with 1\% minimum support and at least 30\% with the \textit{towards closure} outcome} 
	\label{tbl:prediction}
\end{table}

Besides summarizing how companies have behaved so far, the prototype tool also allows the analyst to reason about future outcomes. In the most troublesome event paths in Figure \ref{fig:viz_comp} at least 30 \% of the companies have experienced \textit{risk of bankruptcy}. If an analyst employs this heuristic to narrow down the search for future troublesome companies, i.e. companies going through subpaths where at least 30 \%  have had the negative outcome and with 1 \% minimum support, the resulting subgroups are presented in Table \ref{tbl:prediction} for all three aggregation methods on the three datasets. The table includes both size of the subgroups, \textit{risk of bankruptcy} percentage within the subgroup as well as the chance of picking a company from the current normal companies that will experience \textit{risk of bankruptcy} in the future, i.e. prediction precision for the period 2014-2017. Note that the table does not say anything about prediction recall. Both the MSP and SA methods find larger interesting subgroups compared to the MCP method and they have higher prediction precision. While the subgroups of the MSP method are slightly larger, the SA method have higher prediction precision and thus generalizes better beyond the datasets even though the two methods provided similar information gains in Section \ref{comp_eval}. 

Another interesting observation from Figure \ref{fig:viz_comp} is the growing group of companies starting with the composite event 17, which also can be categorized as a troublesome path. Several of the companies who also goes through this event in the light paths will experience the \textit{risk of bankruptcy} outcome after 2014, i.e. the visualization can also be used to identify interesting subpaths that might hint future outcomes. The composite event 17 can best be described as a major overhaul of the company, since it includes both chairman, board member, business type, name and company form updates as its main components. We will discuss how time related experiments can be incorporated in the user interface in the subsequent section, such that an analyst can better identify which paths are currently the interesting ones. 

% Use time to predict

\section{Discussion} \label{future}
We have shown that combining events into composite events prior to sequence aggregation can provide a better separation of temporal event sequences w.r.t. sequence outcome in the real world domain of business investigation. While the data quality is unknown, the results suggest that the data can still be used for initial separation of the sequences, i.e. the companies, such that guesses about future outcomes are more informed and significantly better than random guessing. Even if an event sequence dataset is not rich enough to provide perfect outcome predictions, the proposed system can show what the most critical flows are and how well they generalize.
% This information can be used to limit the number of companies worth further investigation, but also as input to future learning scenarios. 

Immediate future work includes evaluations and improvements of the user interface with domain experts at the DBA. While the underlying algorithms generalize to multiclass and numeric outcomes, the current visual system is designed for binary outcomes. It is therefore also worthwhile to investigate how the visualizations can be extended to multiclass and numeric outcomes.
We believe the presented system will be very powerful as a complement to existing approaches, since the domain experts currently lack holistic views on the registration data. Furthermore, it will be interesting to either compare the results from this type of analysis with an analysis based on other datasources, e.g., financial statements, or fuse with other datasources in the feature generation for the composite events. 
%While our approach for computing composite events abstracts away minor parts of the sequence ordering, it also provides a way to merge temporal event sequence analysis with other types of data by adding additional features.  

In general, composite event learning opens new possibilities but also poses several challenges. Understanding clustering results is inherently difficult, hence the average user will probably find it difficult to make sense of the high level events that are formed from several event types. Further investigations into how best to convey the components of a composite event are therefore necessary, i.e. the visualizations in Figure \ref{fig:viz_comp}(c). Basic interactions like manual updates to a composite event or labeling of an event with a user-friendly name for future reference could be incorporated in future iterations of the system. Choosing appropriate parameters for the composite event learning, i.e. segmentation window size and number of clusters, is also a difficult task for the average user, thus it becomes important to show users how varying parameters affect both the resulting composite events and the quality metrics. Additionally, automatic suggestions for parameters that score well on the different quality metrics can be included in the user interface to support users that are less familiar with parameter tuning. Furthermore, we want to investigate how best to combine the different aggregations methods with composite event learning. Currently, we compute the full event hierarchy after the sequence simplification, but any of the other methods for extracting patterns can also be used after sequence simplification. This opens for the opportunity to compute multiple composite event candidates and afterwards reason about which ones are appropriate for a given analysis task by, e.g., extracting the most separating pattern and scoring it using the quality metrics. Multiple composite event candidates can, e.g., be based on different feature subsets, in cases where not all event types in a given window should influence the clustering, or a combination of both raw event types and composite events. If event types also have attributes -- as in the domain of business investigation where, e.g., the event \textit{business type change} also includes an attribute with the new business type -- multiple candidates can be computed by replacing an event type with the corresponding attribute, if it is categorical, or include numeric attributes as features in the clustering.  

When event sequences are not only ordered but also timestamped, several choices have to be made regarding proper data extracts for both pattern extraction and evaluation as described in Section \ref{eval}. Future work about how different choices to these parameters can seamlessly be incorporated into the user interface is also interesting, such that history is continuously used to assess the relevance of the computed patterns. This also means that timely insights, i.e. information about absolute time as in Figure \ref{fig:viz_comp}(g), should be incorporated at opportune steps in the overall analysis flow. For instance, in the business investigation domain, using information about when certain updates happened can narrow down the subset of interesting companies even further compared to only using the event tree. We also believe that users should be able to fluently shift between different levels of abstraction, since an analyst then will be able to browse the overall patterns using the composite events to identify interesting subgroups and afterwards easily zoom in on the actual event sequences of the chosen subgroup. 

Effective solutions to much of the future work discussed in this section are most likely domain specific, hence we would also like to apply our methods to different domains. While the exact match between simplification and aggregation method might change from domain to domain, or even from case to case, we believe the approaches presented in this paper generalize to other domains.

% \subsection{Future Work} 
% \subsection{Alternative Domains}

%\subsection{Multiple Composite Event Candidates}

%\subsection{Composite Event Details}

%\subsection{Using History to Assess Relevance} \label{time_choices}

%\subsection{Timely Insights}

%\subsection{Level of Abstraction Interactions}

\section{Conclusion}
In this paper, we present the idea of composite event learning to simplify large collections of temporal event sequences prior to pattern extraction or aggregation. We compare our approach with a recent branching pattern algorithm \cite{liu2017coreflow} that computes the most common event flow as well as a modified version, where we use theory from decision tree construction to compute the most separating pattern w.r.t. sequence outcome. All methods are able to compute event hierarchies that are suitable for visualization. Evaluation results, using relevant pattern quality metrics, show that computing composite events is useful in the real world domain of company investigation and that the unsupervised aggregation method based on composite event learning is better for future outcome prediction compared to the supervised pattern extraction method of the raw event sequences.

We have also designed a visual analytics system prototype that incorporates the simplification and aggregation methods. The goal is to support domain experts at the Danish Business Authority identify critical event flows w.r.t. chosen sequence outcomes, such as bankruptcy. The system allows analysts to efficiently visualize separating flows, and thereby gain insights into how different composite events affect outcome probabilities, as well as inspect the components of the composite events. We also present a use case example that shows how the learning algorithms and visualizations combined can assist the domain expert in gaining relevant insights. Future work include user studies and ways to assist experiments with time in the user interface, which is important when the goal is to reason about the future.

\acknowledgments{
This work was conducted in the DABAI project (IFD-5153-00004B) supported by the Innovation Fund Denmark.}

\bibliographystyle{abbrv-doi}

\bibliography{references}
\end{document}